\documentstyle[preprint,revtex]{aps} 
\input{epsf}

\begin{document}
\pagestyle{myheadings}
\markright{\today}
\def\lapprox{{\raise0.5ex\hbox{$<$}\hskip-0.80em\lower0.5ex\hbox{$\sim$}
}}
\def\gapprox{{\raise0.5ex\hbox{$>$}\hskip-0.80em\lower0.5ex\hbox{$\sim$}
}}

\begin{title}
\begin{center}
{\bf Observation of strong final-state effects in $\pi^+$ production \\ 
in pp collisions  at 400 MeV}
\thanks{supported by the BMBF (06 TU 886), DFG (Mu 705/3,
Graduiertenkolleg), DAAD and NFR and INTAS-RFBR}
\end{center}
\end{title}
\author{A.~Betsch$^1$, R.~Bilger$^1$, W.~Brodowski$^1$, H.~Cal\'en$^{2}$,
H.~Clement$^1$, J.~Dyring$^{2}$,  
C.~Ekstr\"om$^{3}$,  K.~Fransson$^{2}$, L.~Gustafsson$^{2}$,
S.~H\"aggstr\"om$^{2}$, B.~H\"oistad$^{2}$, J.~Johanson$^{2}$,
A.~Johansson$^{2}$, 
T.~Johansson$^{2}$, K.~Kilian$^{4}$, S.~Kullander$^{2}$,
A.~Kup\'s\'c$^{5}$, G.~Kurz\thanks{Present address: ETH Z\"urich/PSI Villigen,
  Switzerland}, 
P.~Marciniewski$^{5}$, B.~Morosov$^{6}$,
A.~M\"ortsell$^{2}$, W.~Oelert$^{4}$,
R.J.M.Y.~Ruber$^{2}$, M.G.~Schepkin$^7$,  J.~Stepaniak$^{5}$,
A.~Sukhanov$^{6}$, A.~Turowiecki$^{8}$, G.J.~Wagner$^1$, Z.~Wilhelmi$^{8}$,
J.~Zabierowski$^{9}$, A.~Zernov$^{6}$, J.~Z\l oma\'nczuk$^{2}$}
\begin{instit}
$^1$Physikalisches Institut, Universit\"at T\"ubingen, 72076
T\"ubingen, Germany \\
$^2$Dept of Radiation Sciences, University of Uppsala, Sweden \\
$^{3}$The Svedberg Laboratory, Uppsala, Sweden \\
$^{4}$Institut f\"ur Kernphysik, Forschungszentrum J\"ulich, Germany \\
$^{5}$So\l tan Institute for Nuclear Studies, Warsaw, Poland \\
$^{6}$Joint Institute for Nuclear Research, Dubna, Russia \\
$^7$Institute for Theoretical and Experimental Physics, Moscow \\
$^{8}$Institute of Experimental Physics, Warsaw University,  Poland \\
$^{9}$Institute for Nuclear Studies, L\'odz, Poland 
\end{instit}

\begin{abstract}
Differential cross sections of the reactions $pp \to d\pi^+$ and $pp \to
pn\pi^+$ have been measured at $T_p = 400$ MeV by detecting the charged
ejectiles in the angular range $4^0 \leq \Theta_{Lab} \leq 21^\circ$. The deduced
total cross sections agree well with those published previously for neighbouring
energies. The invariant mass spectra  are observed to
be strongly affected by $\Delta$ production and $NN$ final-state interaction.
The data are well described by Monte Carlo simulations including both these
effects.  The ratio of  $pp \to pn\pi^+$ and $pp \to d\pi^+$ cross sections
also compares favourably to a recent theoretical prediction which suggests a
dominance of $np$-production in the relative $^3S_1$-state.
\end{abstract}

\vspace{\bigskipamount}

\narrowtext

The single pion production has received renewed interest in recent years both
from experimental and theoretical points of view. Theoretical aspects
currently under discussion include possible heavy meson exchange, the nature
of the $\pi NN$ vertex and the role of final-state interactions (FSI) in these
reactions. Experimentally the availability of storage rings has opened the
possibility to measure single pion production with high statistics even close
to threshold \cite{har97,fla98,bon95,mey90}. These  data show an energy
dependence of the total cross section near threshold, which deviates
substantially from phase space suggesting a strong influence of the  $NN$ FSI.
In this Letter we show that  FSI effects can be 
explicitely observed and identified  in the  invariant $(M)$ and
missing mass $(MM)$ spectra of the reaction $pp \to pn\pi^+$. The energy of 
$T_p = 400$  MeV is already high enough for $\Delta$-production to be observed
clearly in $M_{n\pi^+}$ and $M_{p\pi^+}$, whereas the $pn$ FSI still strongly
influences $M_{pn}$ even though this energy is well above threshold.

The measurements have been performed at the CELSIUS storage ring at
$T_p = 400$~MeV using the WASA/PROMICE detector setup including a
hydrogen cluster jet target. Details of the detector and its
performance are given in \cite{cal96}. For the data presented here
only the  forward detector has been utilized, which allows
the determination of the four-momentum of charged particles in the
angular range of $4^\circ \leq \Theta_{Lab} \leq 21^\circ$. It
is composed of a tracker with  proportional
counter straw chambers for an accurate determination of particle
trajectories, followed by segmented scintillator trigger and range
hodoscopes for dE and E determinations, respectively. Particle
identification has been made by use of the dE-E method.
Fig.~\ref{fig1} shows a three-dimensional plot of the dE-E spectrum.
Deuterons, protons and pions appear well separated. For the $\pi^+$
identification the delayed pulse technique (observation of the delayed
pulse from $\mu^+$ decay following the $\pi^+$ decay) has been
utilized in addition. Though protons and deuterons appear to be well
separated in Fig.~\ref{fig1}, their separation is not perfect and the
contamination of reconstructed $pn\pi^+$ events with $d\pi^+$ events,
and vice versa, is non-negligible. In order to get rid of this
contamination we have requested the $d\pi^+$ events to be planar
$(\Delta \Phi = 170^\circ - 190^\circ)$ and the $p\pi^+$ events to be
non-planar. This way mutual contaminations could be kept below 1\%. The final
number of good events which passed all criteria has been about $10^5$ for
$d\pi^+$ and $7 \ast 10^4$ for $pn\pi^+$.
The integral luminosity has been determined to better than 5\% by the
simultaneous measurement of $pp$ elastic scattering and its comparison
to literature values \cite{cha97}. Detector response and efficiencies
have been determined by Monte-Carlo (MC) simulations, utilizing the
program package GEANT \cite{geant} including the treatment of secondary
interactions in the detector.

The data obtained for $pp \to d\pi^+$ are shown in Fig.~\ref{fig2}. In
the upper part the $\pi^+$ missing mass $MM_{\pi^+}$ which gives a
peak at the position of the deuteron mass is displayed together with
the corresponding MC simulations. The good agreement
demonstrates that the detector response is understood to high
precision.  The measured $\pi^+$ angular distribution in the center-of-mass
(c.m.)  system
is displayed in the lower part of Fig.~\ref{fig2} together with the
SAID phase shift prediction \cite{cha97}. Note that the angular
distribution is symmetric about $90^\circ$ due to the symmetry in the
entrance channel. The limited angular range of the data results from
the experimental requirements of $4^\circ \leq \Theta_{Lab} \leq
21^\circ$ for both $d$ and $\pi^+$. The experimental points are 
 compatible with  {SAID}, and we may use the angular
dependence of the latter to extrapolate our data to $4\pi$.
This way we obtain a total cross section  $\sigma (pp \to d\pi^+) = 0.78$~mb,
which 
compares quite favorably with the SAID value of 0.82 mb. Whereas the
statistical uncertainty is less than 1\%, the systematic uncertainty
is estimated to be about 7\% comprising uncertainties both from the
determination of the luminosity and from the handling of deuteron
breakup in the detector by the MC simulation.

Results for the $pp \to pn\pi^+$ reaction are comprised in
Figs.~\ref{fig3}--\ref{fig5} . The invariant and missing mass spectra
reconstructed from the measured $p\pi^+$ events are shown in
Fig.~\ref{fig3} together with curves from MC simulations assuming pure phase space
(dotted) and including either $\Delta^{++}(\Delta^+)$ excitation in
the $p\pi^+(n\pi^+)$ system (dash-dotted) or the s-wave FSI in the
$pn$ system (dashed). For the latter the effective range approximation
of Migdal-Watson type as given in \cite{sch93} has been used. The only
free parameter, $R$, accounts for the size of the interaction region,
from which the two nucleons emerge, and effectively subsumes also the
part of the reaction, where the two nucleons are not in relative
s-waves. For a point-like vertex one finds $R \approx 0.8$ fm
\cite{sch93}. Here in the MC simulations $R$ has been adjusted for
best reproduction of the data resulting in $R = 2.5$ fm, if we assume
the s-wave $pn$ system to be in pure $^3S_1$ states as previous work
suggests \cite{bou96,fal85}. This assumption will also be corroborated by a
comparison of the $d\pi^+$ and $pn\pi^+$ channels below. The value for $R$ is
compatible with those obtained in analyses of the $NN$ FSI
in the deuteron breakup on the proton at low energies \cite{bru69}.
The $\Delta$ excitation in the exit channel is assumed to take place
between the $\pi^+$ and either the proton or the neutron according to
the isospin ratio of $9 : 1$ with the $\pi^+$ being in relative
p-state in the $\pi N$ system. Note that parity conservation requires the pion
also to be in a $p$-wave relative to  $^3S_1~pn$ final state.

The MC simulation including both FSI effects together is displayed in
Fig.~\ref{fig3} by the shaded histograms which reproduce the data very
well. The invariant mass spectrum $M_{p\pi^+}$ and the missing mass
spectrum $MM_p$ (corresponding to $M_{n\pi^+}$) are peaked towards
large masses, whereas $MM_{\pi^+}$( corresponding to $M_{pn}$) is
strongly peaked towards low masses in contrast to phase space
distributions (dotted lines). In all three spectra both FSI effects
play a significant role. However, the fact that the peaking towards
large masses is much more pronounced in $M_{p\pi^+}$ than in $MM_p$,
clearly exhibits a strong influence of the $\Delta$ production, since
it is nine times more likely in the $p\pi^+$ system than in the
$n\pi^+$ system. This observation in $M_{p\pi^+}$ and $M_{n\pi^+}$ is
strongly different from the one at $T_p = 310$ MeV, where both spectra
have been observed to be of comparable shape \cite{har97}. However, it
is in agreement with the trend observed for $T_p \leq 330$ MeV
\cite{fla98}. From the partial-wave analysis of those data it was
found \cite{fla98} that the resonant p-wave contribution, though still
small at these energies, is steeply rising with incident energy. From
this analysis the $\Delta$ contribution is expected to get dominant at
energies of $T_p \approx 400$ MeV (see Fig.~20 of Ref. \cite{fla98}).
This is indeed what we observe in our data.  We note  that
the strong influence of the FSI on the pion energy distribution is  already
apparent in the dE-E plot in Fig.~\ref{fig1}, where the pions of the
$pn\pi^+$ reaction are seen to be concentrated towards the peak of the
$d\pi^+$ reaction.

In Fig.~\ref{fig4} the experimental pion and proton angular distributions,
corrected for detector  efficiency and acceptance, 
are compared to MC simulations for the reaction $pp \to pn\pi^+$ . As in 
$pp \to d\pi^+$ the angular distributions have to be symmetric about
$90^\circ$. The proton angular distribution is essentially isotropic
being affected only slightly by the $\Delta$ excitation. The pion
angular distribution, on the other hand, depends strongly on
the $\Delta$ excitation. Conventionally the pion angular distributions
in single pion production are parametrized by $\sigma(\Theta_\pi) \sim
1/3 + b \cos^2\Theta_\pi$, where $b$ is the so-called anisotropy
parameter. Previous analyses yielded values for $b$ up to 0.5 for $np
\to nn\pi^+$ \cite{ban94} and 0.3 for $pp \to pp\pi^0$ \cite{rap95}
with the maximum $b$ being reached near $T_p \approx 550$ MeV. 
The $b$ values are observed to decrease with decreasing
$T_p$. At $T_p = 460$ MeV, the lowest energy analyzed in those studies,
$b$ gets as small as 0.1. On the other hand the IUCF measurements
\cite{har97,fla98} of $pp \to pn\pi^+$ show a strong rise of $b$
already close to threshold  reaching $b = 0.23(6)$ at their
highest  energy of $T_p = 330$ MeV. Whereas from the latter
we would expect to find already a quite substantial value of $b$ at
400 MeV, the previous analyses would suggest rather a small value.
Hence we show in Fig.~\ref{fig4} two MC calculations including $pn$
FSI and $\Delta$ excitation, one with $b = 0.1$ (dashed lines) and one
with $b = 0.4$ (solid)  --- in addition to the phase-space expectation (dotted
line). In the measured range of $\Theta_\pi$ the data
clearly prefer the larger $b$ value. From the $\cos^2(\Theta_\pi)$ plot of
the pion angular distribution in Fig.~\ref{fig4} it is readily seen
that $b=0.7$ would be an upper limit  compatible with our
data.  For a more precise determination of $b$  larger pion
angles are necessary, which are not covered by this measurement. Hence also the
determination of the total cross section from our data is not
independent of the assumption for $b$. Whereas for $b = 0.1$ one would
obtain a value of $\sigma(pp \to pn\pi^+) = 0.73(4)$~mb, our data favour
values of 0.62(4) and 0.57(4) mb for $b = 0.4$ and 0.7, respectively,
where the assigned uncertainty is essentially due to that of the
luminosity. The latter values are in good agreement with literature
values at neighbouring energies (see \cite{har97}).

The close relation between $d\pi^+$ channel and $pn\pi^+$ channel as the
breakup channel of the former has recently been pointed out by Boudard, F\"aldt
and Wilkin \cite{bou96}. Assuming that the final $pn$ system is in the $^3S_1$
state, and that the pion production operator is of short range,
they derive  a simple relation
(eq. 6 of ref. \cite{bou96}) for the ratio of differential cross sections
defined by 
$$R(Q) = 2\pi B_d {d^2\sigma \over d\Omega_\pi dQ} (pp \to
pn\pi^+)/{d\sigma \over d\Omega_\pi} (pp \to d\pi^+) = {p(x) \over
  p(-1)} {\sqrt{x} \over 1 + x}$$ 
which should be independent of the
pion scattering angle. Here
$B_d$ denotes the deuteron binding energy and $Q = M_{pn} - m_d$ the
excitation energy in the $pn$ system, $x = Q/B_d$ and $p(x)$ is the
pion c.m. momentum. Our experimental result for $R(Q)$ is plotted in
Fig.~\ref{fig5} together with the prediction of Ref. \cite{bou96}.
Though there is good qualitative agreement, the data exhibit a
significantly steeper slope than the prediction. In particular, the
data are higher at low $Q$, where the approximations made in Ref.
\cite{bou96} should be valid best. Experimentally the determination of
$R(Q)$ from the simultaneous measurement of both reactions is expected
to be particularly reliable, since uncertainties in the determination
of the luminosity and detector response cancel to large degree in the
ratio, the only major source of possible error being the treatment of
the deuteron breakup in the detector.
However, since our result for $\sigma(pp \to d\pi^+)$ agrees well
with the literature values at neighbouring energies, we do
not see a significant problem there either. A closer inspection of
Fig.~1 in ref. \cite{bou96} indicates that experimentally $R(Q)$ is
not fully angle independent. The TRIUMF data \cite{fal85} plotted
there exhibit a systematic trend around the maximum of $R(Q)$: the
data taken at small angles $(\Theta_{Lab} =46^\circ, 56^\circ)$ lie
systematically above the theoretical curve, whereas the ones taken at
larger angles $(73^\circ - 88^\circ)$ lie significantly below. Our
result for the forward angular range fits very well into this trend in
the TRIUMF data. Since in the calculation of $R(Q)$ only the isoscalar
$^3S_1$ channel of the pn system is considered, the surplus of
approximately 10\% in our data at small $Q$ may be associated with
contributions from other partial wave channels, notably the $^1S_0$
channel of the pn system. This conclusion conforms quite nicely with
the expectations from the partial-wave analysis at lower energies
\cite{fla98}. The observed discrepancy between calculation and data at
higher $Q$ may be associated with effects from higher partial waves as
well as with kinematic approximations made in the derivation of the
theoretical expression for $R(Q)$ \cite{bou96}.

Summarizing we observe in the exclusive measurement of $pp \to
pn\pi^+$ at $T_p = 400$ MeV strong FSI effects in the invariant and
missing mass spectra which are identified as being  due to $pn$ FSI and 
$\pi N\Delta$ excitation. 
The dominance of the latter is in agreement with expectations from
partial-wave analyses close to threshold. The influence of the former
is in agreement with expectations for the Migdal-Watson effect in the
$pn$ system, if s-waves are dominating there. The observed ratio
$R(Q)$ for the exit channels $pn\pi^+$ and $d\pi^+$ is compared 
with the prediction of ref. \cite{bou96} relating the channel of the
bound system with its breakup channel. Although the general agreement is good,
there are 
significant differences  which call for
a more refined theoretical treatment of both channels.

We gratefully acknowledge valuable discussions with Colin Wilkin. We are also
grateful to the personnel at the The Svedberg Laboratory for their help during
the course of this work.

\figure{Plot of the dE-E spectrum for two charged particles in the forward
  detector (FD). The energy loss dE has been measured by the forward trigger
  hodoscope (FTH), see ref. \cite{cal96}. Energies are given in units of GeV.
\label{fig1}}

\figure{Top: Spectrum of the $\pi^+$ missing mass $MM_{\pi^+}$ as obtained from
  identified $d\pi^+$ events together with the corresponding MC simulation
  (shaded area) of the detector response. Bottom: Measured $\pi^+$ angular
  distribution for $pp \to d\pi^+$ in comparison with the SAID phase shift
  prediction  \cite{cha97}.
\label{fig2}}

\figure{Spectra for $p\pi^+$ invariant mass $M_{p\pi^+}$
  and missing masses $MM_p$ and $MM_{\pi^+}$ for $pp \to pn\pi^+$
  reconstructed from the $p\pi^+$ events detected for 
  $4^\circ \leq \Theta_{Lab} \leq 21^\circ$. Corresponding MC simulations are
  shown assuming pure 
  phase space (dotted), including either $pn$ FSI (dashed) or $\Delta$
  excitation (dash-dotted) and both together (shaded histograms). 
\label{fig3}}

\figure{Angular distributions of protons (top) and pions (bottom) for
  $pp \to pn\pi^+$ reconstructed from the detected $p\pi^+$ events.
  They are plotted in dependence of $\cos^2(\Theta)$ and are compared
  to MC simulations assuming pure phase space (dotted) and including
  both $pn$ FSI and $\Delta$ excitation with $b = 0.1$ (dashed) or $b
  = 0.4$ (solid). All simulations are normalized to an integral cross
  section of $\sigma=0.62$~mb.
\label{fig4}}

\figure{Ratio $R(Q)$ of the $pp \to pn\pi^+$ and $pp \to d\pi^+$ differential
  cross sections. The
  solid curve shows the prediction of Ref. \cite{bou96}. 
\label{fig5}}

\newpage

\begin{center}
\epsfxsize=0.8 \textwidth
\mbox{\epsfbox{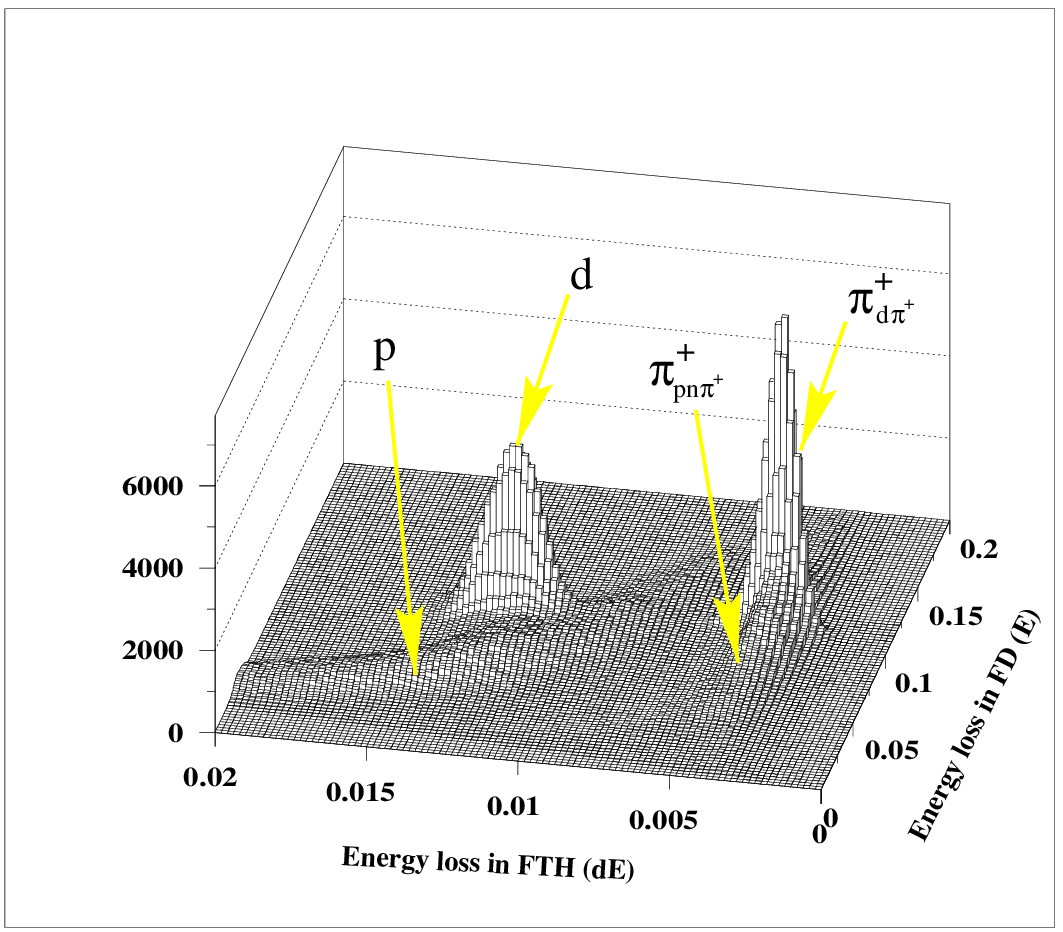}}\\
Fig.~1
\end{center}

\begin{center}
\epsfxsize=0.8 \textwidth
\mbox{\epsfbox[38 360 512 1104]{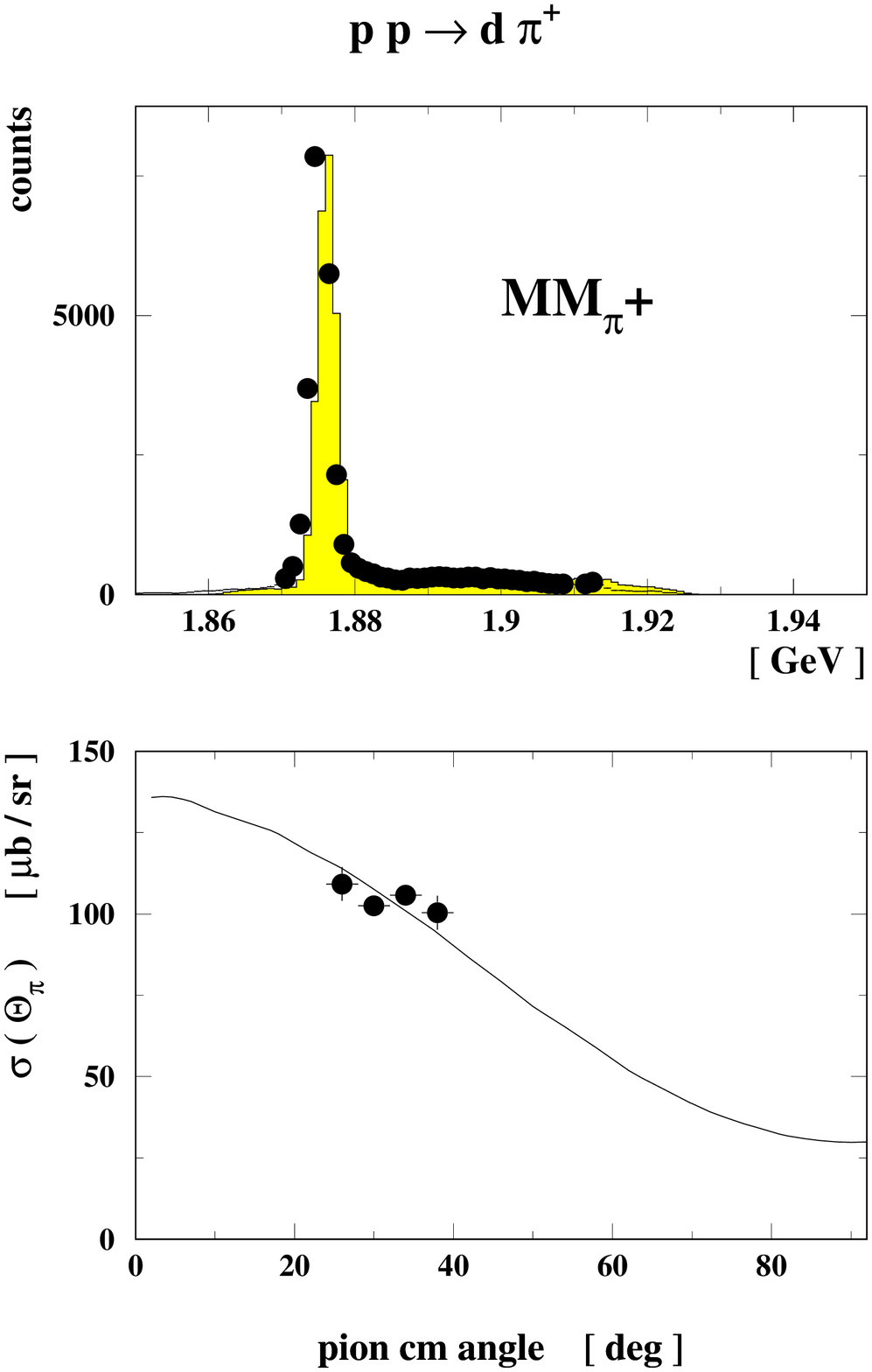}}\\
Fig.~2
\end{center}

\begin{center}
\epsfxsize=0.65 \textwidth
\mbox{\epsfbox{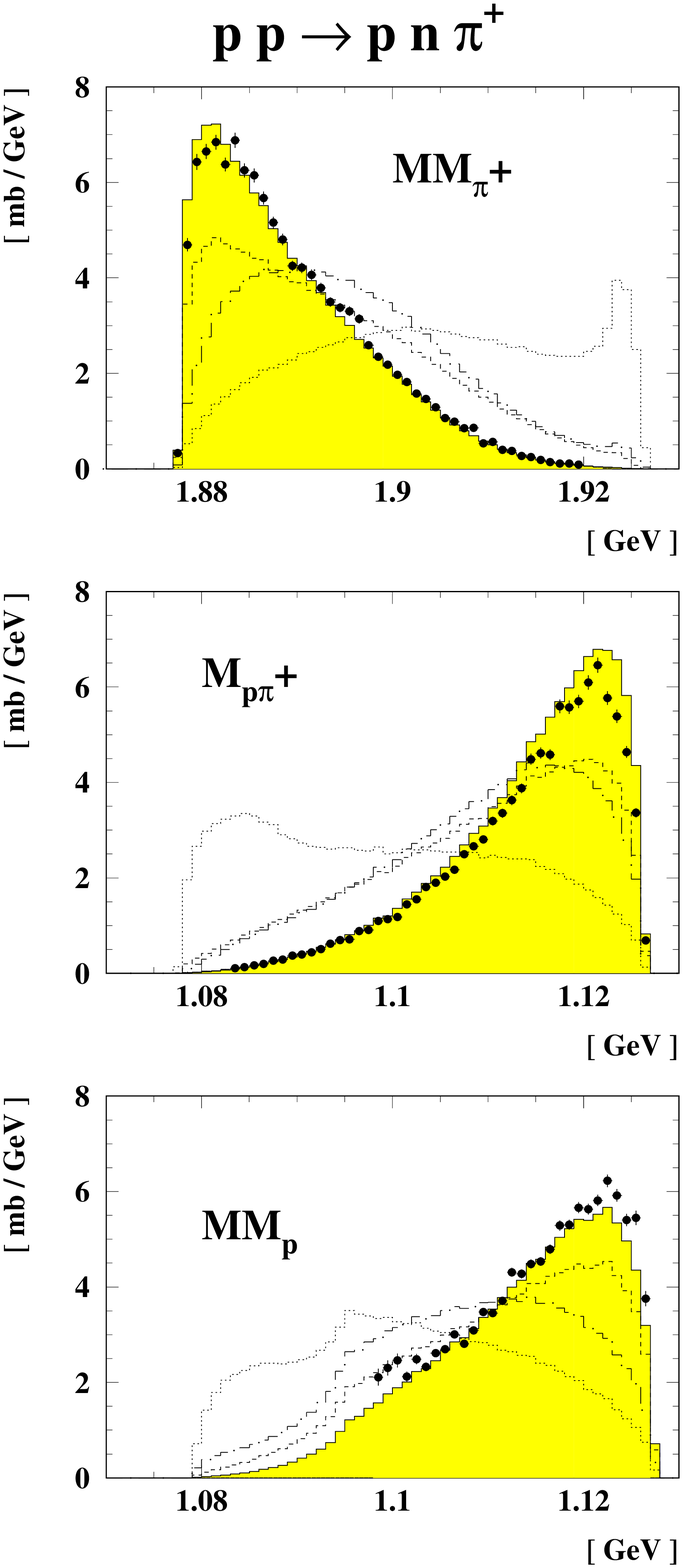}}\\
Fig.~3
\end{center}

\begin{center}
\epsfxsize=0.8 \textwidth
\mbox{\epsfbox[30 366 526 1108]{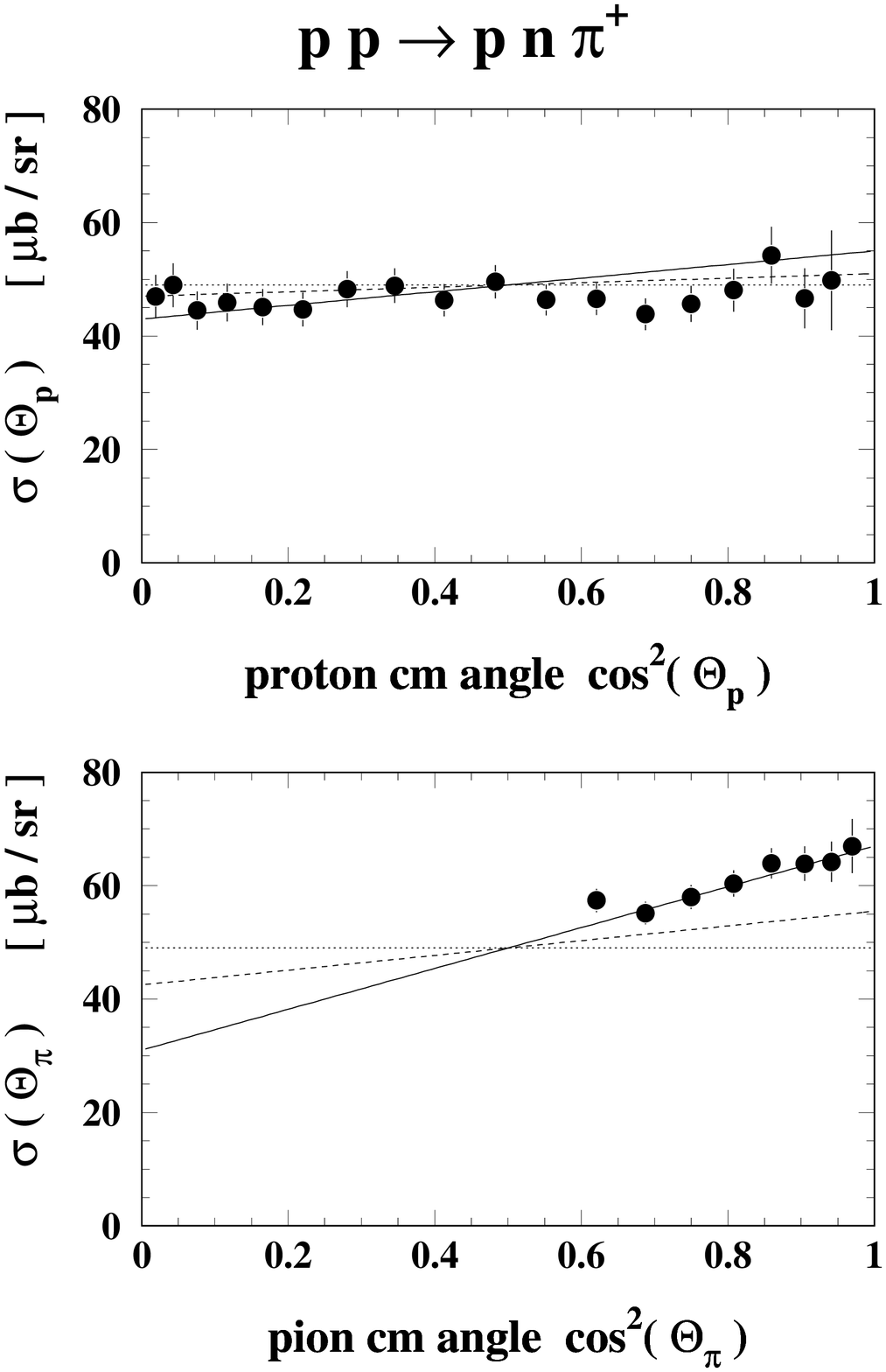}}\\
Fig.~4
\end{center}

\begin{center}
\epsfxsize=0.8 \textwidth
\mbox{\epsfbox[32 522 380 778]{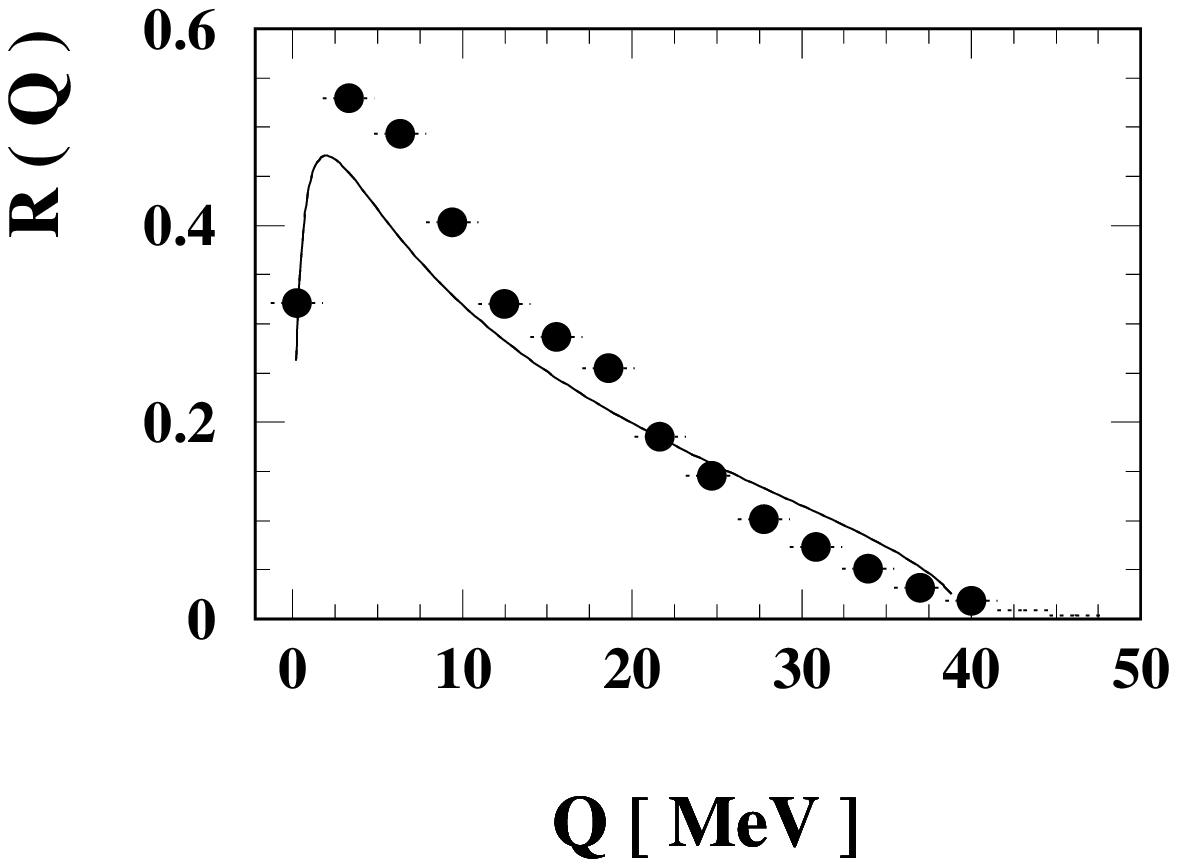}}\\
Fig.~5
\end{center}

\end{document}